\documentclass[a4paper,fleqn]{cas-dc}

\usepackage[numbers]{natbib}
\usepackage{float}
\usepackage{algorithm}
\usepackage{algpseudocode}
\setcitestyle{square}
\usepackage{appendix}
\usepackage{soul,xcolor}
\usepackage{booktabs, makecell}

\newlength\myindent
\def\tsc#1{\csdef{#1}{\textsc{\lowercase{#1}}\xspace}}
\tsc{WGM}
\tsc{QE}

\begin{document}
\let\WriteBookmarks\relax
\def\floatpagepagefraction{1}
\def\textpagefraction{.001}

\shorttitle{Exploring the impact of noise and degradations on heart sound classification models}    

\shortauthors{D Shariat Panah, A Hines, S McKeever}  

\title [mode = title]{Exploring the impact of noise and degradations on heart sound classification models}  



%

\author[1]{Davoud Shariat Panah}[orcid=0000-0003-1940-9968]

\cormark[1]


\ead{davoud.x.shariatpanah@mytudublin.ie}



\affiliation[1]{organization={School of Computer Science, Technological University Dublin},
            city={Dublin},
            country={Ireland}}

\author[2]{Andrew Hines}[]


\ead{andrew.hines@ucd.ie}



\affiliation[2]{organization={School of Computer Science, University College Dublin},
            city={Dublin},
            country={Ireland}}

\author[1]{Susan McKeever}[]


\ead{susan.mckeever@tudublin.ie}



\cortext[1]{Corresponding author}



\begin{abstract}
The development of data-driven heart sound classification models has been an active area of research in recent years. To develop such data-driven models in the first place, heart sound signals need to be captured using a signal acquisition device. However, it is almost impossible to capture noise-free heart sound signals due to the presence of internal and external noises in most situations. Such noises and degradations in heart sound signals can potentially reduce the accuracy of data-driven classification models. Although different techniques have been proposed in the literature to address the noise issue, how and to what extent different noise and degradations in heart sound signals impact the accuracy of data-driven classification models remains unexplored. To answer this question, we produced a synthetic heart sound dataset including normal and abnormal heart sounds contaminated with a large variety of noise and degradations. We used this dataset to investigate the impact of noise and degradation in heart sound recordings on the performance of different classification models. The results show different noises and degradations affect the performance of heart sound classification models to a different extent; some are more problematic for classification models, and others are less destructive. Comparing the findings of this study with the results of a survey we previously carried out with a group of clinicians shows noise and degradations that are more detrimental to classification models are also more disruptive to accurate auscultation. The findings of this study can be leveraged to develop targeted heart sound quality enhancement approaches -  which adapt the type and aggressiveness of quality enhancement based on the characteristics of noise and degradation in heart sound signals.
\end{abstract}



\begin{keywords}
 Phonocardiogram \sep
 Heart sound \sep Heart sound classification \sep Noise and degradation \sep  Quality enhancement \sep Synthetic dataset
\end{keywords}

\maketitle

\section{Introduction}\label{sec:intro}

Cardiovascular diseases are currently the leading cause of mortality worldwide, accounting for one-third of deaths globally \cite{Cardiovascular}. Early diagnosis through pervasive approaches can help detect heart disease in patients at earlier stages and consequently improve the survival rate. Auscultation has been a cost-effective approach for pre-screening heart disease for over 200 years \cite{Shaver1995Cardiac}. However, auscultation is a subjective practice and requires extensive training \cite{2015Learning}. Therefore, automatic analysis of heart sounds has been presented as an alternative to auscultation for early pre-screening of heart abnormalities. In recent years, a large variety of data-driven heart disease diagnostic systems have been developed that can distinguish normal from abnormal heart sounds \cite{Gupta2007Neural,Abduh2020Classification,Alkhodari2021Convolutional,Dhar2021Cross-wavelet,Er2021Heart}.

Heart sounds are generally captured using digital stethoscopes or mobile phones. Although such devices typically benefit from noise reduction and cancellation technologies \cite{3M,CORE,Jabes}, they can still capture a considerable amount of noise while recording heart sounds, especially in noisy environments. Due to the presence of internal physiological body noises and ambient artefacts in clinical and non-clinical settings, it is almost impossible to record noise-free heart sound signals in real-world scenarios.

It has been stated that noise and contaminations in heart sound recordings can reduce the performance of data-driven models \cite{Paul2006Noise,Kumar2011Noise,Gradolewski2014Wavelet-based,Jain2017adaptive}. Researchers have adopted different approaches to mitigate the negative impact of noise and degradations in captured signals on the performance of data-driven heart disease diagnostic systems. Heart sound quality enhancement is one of the most widely adopted approaches to reducing noise in captured signals. A whole host of enhancement techniques have been employed in the field, such as filtering \cite{Gupta2007Neural,Abduh2020Classification} and wavelet-based denoising \cite{Gradolewski2014Wavelet-based,Jain2017adaptive,Chen2020Classification,Messer2001Optimal}. Heart sound quality classification is another approach in which a data-driven model distinguishes low-quality heart sounds from good-quality ones \cite{Springer2016Automated,Naseri2012Computerized,Mubarak2018Analysis}. Low-quality signals are discarded, while heart sounds with an acceptable quality will subsequently be used in heart disease diagnostic models.

While numerous methods have been proposed in the field to enhance or classify the quality of heart sounds, it has remained unexplored how and to what extent noise and degradation in heart sound signals can impact the overall accuracy of  data-driven models. A deeper understanding of the impact of noise and degradations in heart sound recordings on the performance of the data-driven models will allow us to adapt the heart sound capture process with the aim of minimizing the negative impact of such noise and degradations. Also, such an understanding will enable us to adjust the quality enhancement of the captured heart sounds based on their noise content and develop targeted heart sound pre-processing pipelines. In this regard, this study aims to answer the following research question: how do noise and degradations in heart sound signals impact the overall accuracy of data-driven models?

To answer this research question, we will produce a synthetic dataset containing normal and abnormal heart sounds with various noises and degradations. This dataset will then be employed to train and evaluate multiple heart sound classification models. This will enable us to systematically investigate the impact of a large variety of noises and degradations on the performance of different heart sound classification models. Previously we investigated the impact of noise and degradations on heart sound signals' diagnosability by conducting a survey with a group of trained clinicians \cite{ShariatPanah2022audio}. In this study, we will also observe the similarities between the impact of noise and degradations on the performance of classification models and the results of our previous survey.

The remainder of this paper is structured as follows: Section \ref{sec:rw} overviews the related work on the impact of noise and the application of synthetic datasets. Section \ref{sec:methods} provides the details of the datasets and data-driven models employed in this study. In Section \ref{sec:result}, the results are given. In Section \ref{sec:disc}, results are discussed. Conclusions and future directions are presented in Section \ref{sec:conc}.

\section{Related Work}\label{sec:rw}

\subsection{Noise Impact}\label{sec:rw-ni}

Noise and degradations with internal or external sources can have a negative impact on auscultation. Shindler \cite{Shindler2007Practical} has stated that high levels of environmental sounds, such as speech, can interfere significantly with auscultation. Coviello \cite{Coviello2013Auscultation} has pointed out that noises due to muscular movements can interfere with heart sounds and make it harder for clinicians to perceive the salient characteristics of the heart sounds. In addition to ambient and movement noises, internal physiological noises can also be disruptive to auscultation. Ranganathan et al. \cite{Ranganathan2015art} have emphasized that intense breathing noises can interfere with assessing heart sounds and reduce the accuracy of auscultation. In our previous work \cite{ShariatPanah2022audio}, we investigated the impact of noise and degradations on the diagnosability of heart sound recordings through a survey with clinicians. The results of this survey showed that clinicians found ambient noises more disruptive than movement or internal noises. Also, they indicated that long-duration noises are more problematic to accurate auscultation than short-duration noises.

Noise and degradation in heart sound recordings can also reduce the accuracy of data-driven classification models. Paul et al. \cite{Paul2006Noise} have indicated that internal or external noises can mask fundamental heart sounds and increase the false positives of the classifiers. According to Kumar et al. \cite{Kumar2011Noise}, noise in heart sound recordings can alter the morphological characteristics of the heart sounds and change the features salient to accurate diagnostics. Gradolewski et al. \cite{Gradolewski2014Wavelet-based} have stated that some heart sounds, such as late-systolic and pan-systolic murmurs, have similar characteristics to noise, and, as a result, applying denoising algorithms can decrease the misclassification of such signals by data-driven models. Jain et al. \cite{Jain2017adaptive} have indicated that noise and degradations in heart sound signals can reduce the accuracy of the segmentation of heart sounds into heartbeat cycles, which in turn can lead to a sub-optimal heart sound classification model. Although it has been emphasized that noise and degradation in heart sound signals can potentially reduce the accuracy of classification models, we could not find any comprehensive study exploring the impact of different noises and degradations on classification models.

\subsection{Diagnosability of Heart Sounds}\label{sec:rw-dhs}
In our previous work, we surveyed a group of thirteen clinicians to understand the characteristics of diagnosable heart sounds and the impact of noises and degradations on the diagnosability of heart sound recordings \cite{ShariatPanah2022audio}. This survey included a subjective listening test with 20 heart sound recordings contaminated with different noises and degradations. Analyzing the results of this survey showed that noise and degradations in heart sound signals have a detrimental effect on the diagnosability of heart sounds. Most of the survey's respondents (92\%) stated that a diagnosable heart sound recording must contain at least six heartbeat cycles. In addition, the survey results showed that clinicians found ambient noises more disruptive than movement or internal body noises. Also, they indicated that continuous long-duration noises have a more destructive impact on the diagnosability of heart sound recordings than transient short-duration noises.

\subsection{Synthetic Data}\label{sec:rw-sd}
Synthetic data have been widely employed to develop and evaluate data-driven models in different domains. According to Jordon et al. \cite{Jordon2020Synthetic}, synthetic data refers to “data generated from a set of easy-to-specify distributions that is used to validate a machine learning model”. Synthetic data can have different use cases. Lack of access to large, annotated datasets is still a challenge in many fields. As a result, in some studies, synthetic datasets were generated to increase the number of available samples used to develop data-driven models. Li et al. \cite{li2018training} produced a synthetic dataset by augmenting a natural speech dataset with synthetic speech. They used this dataset to train a neural speech recognition model. Severini et al. \cite{severini2019automatic} generated a synthetic dataset of cry sounds through acoustic scene simulation. They used this dataset to train a data-driven model for the automatic detection of cry sounds in neonatal intensive care units. Synthetic datasets have also been used to produce generalizable data-driven models in cases where real-world datasets are biased or lack diversity. For instance, Kortylewski et al. \cite{Kortylewski2019Analyzing} generated a synthetic dataset of facial images that includes a much larger variety of parameters, such as lights and head poses, compared to available real-world datasets. Synthetic data has also been used to address class imbalance problem in some datasets, which can lead to sub-optimal data-driven models. For example, Perez-Porras et al. \cite{Perez-Porras2021Machine} have developed different techniques to balance the available samples by generating synthetic data for wildfire prediction. Synthetic data generation is also a common practice in situations where sharing real-world data is severely limited due to legal considerations \cite{Assefa2020Generating} or privacy issues \cite{Surendra2017review}. Synthetic data also allows us to evaluate different methods and algorithms in a controlled setting. In this regard, another important application of synthetic data is in cases where we need to study the impact of different variations in input data on the performance of algorithms. For instance, Bolon-Canedo et al. \cite{Bolon-Canedo2013review} have used eleven synthetic datasets to compare the performance of different feature selection methods. These synthetic datasets vary in terms of non-linearity, noise level, number of irrelevant and redundant features, etc. Having control over these variables in generated datasets enabled them to study their impact on the performance of feature selection algorithms more precisely and identify the advantages and disadvantages of each algorithm.

\section{Method}\label{sec:methods}
In this section, we provide the details of the methodology for this study. The overall approach can be summarized as follows: 
\begin{itemize}
\item A synthetic heart sound dataset including normal and abnormal heart sounds contaminated with a large variety of noise and degradations will be generated. To generate this dataset, clean heart sounds will be mixed with noises of different types, durations and groupings in different SNR levels.
\item The synthetic dataset will be split into train and test sets.
\item Multiple classification models will be developed. To develop these models, different feature representations (log-spectrogram and mel-spectrogram) and two commonly used classifiers in heart classification (support vector machine and convolutional neural network) will be employed.
\item Support vector machine models will be trained using the synthetic training set. Convolutional neural network models will be pre-trained using a dataset called PhysioNet and then will be fine-tuned using the synthetic training set.
\item After training the models, we will use the synthetic test set to evaluate the classification models.
\item To investigate the impact of noise and degradations in heart sound signals on the performance of classification models, we will report the overall accuracies of the models across heart sounds contaminated with different noise types, durations, groupings and SNR levels.
\end{itemize}

In Section \ref{sec:datasets}, we describe the process of generating the synthetic heart sound dataset. Also, we provide the details of another heart sound dataset called PhysioNet used in our experiments. Afterwards, in Section \ref{sec:models}, we explain the stages of developing heart sound classification models, including pre-processing, feature extraction and classification.

\begin{table}[cols=3]
\caption{Details of the clean heart sound recordings used to generate the synthetic dataset}\label{tb:clean_hs}
\begin{tabular*}{\tblwidth}{@{}llr@{}}
\toprule
Recording \# & Type & Duration (s)  \\
\midrule
1            & Normal           & 12.0                      \\
2            & Normal           & 10.1                      \\
3            & Normal           & 13.8                      \\
4            & Normal           & 10.0                      \\
5            & Normal           & 3.0                      \\
6            & Normal           & 12.8                      \\
7            & Normal           & 10.2                      \\
8            & Normal           & 15.0                      \\
9            & Abnormal - Aortic regurgitation           & 12.0                      \\
10            & Abnormal - Aortic stenosis           & 10.9                      \\
11            & Abnormal - Mitral regurgitation           & 12.0                      \\
12            & Abnormal - Mitral stenosis           & 11.2                      \\
13            & Abnormal - Mitral valve prolapse           & 11.5                      \\
14            & Abnormal - Mitral valve prolapse           & 2.5                      \\
15            & Abnormal - S3           & 10.1                      \\
16            & Abnormal - S4           & 10.0                      \\

\end{tabular*}
\end{table}

\begin{table}[cols=3]
\caption{Details of the noise types, their groupings and durations that have been mixed with clean heart sound recordings to generate the synthetic dataset}\label{tb:noise_types}
\begin{tabular*}{\tblwidth}{@{}lccc@{}}
\toprule
Noise Type & Noise Grouping & Noise Duration  \\
\midrule
White            & Color           & Long                      \\
Pink            & Color           & Long                      \\
Red            & Color           & Long                      \\
Sensor movement            & Movement           & Short                      \\
Body movement            & Movement           & Short                      \\
Deep breathing            & Internal           & Long                      \\
Fast breathing            & Internal           & Long                      \\
Coughing            & Internal           & Short                      \\
Digestive sound            & Internal           & Short                      \\
Talking            & Ambient           & Long                      \\
Door open/close            & Ambient           & Short                      \\
Phone ringing            & Ambient           & Long                      \\
Music            & Ambient           & Long                      \\
Water flow            & Ambient           & Long                      \\
TV            & Ambient           & Long                      \\
Dishwasher            & Ambient           & Long                      \\
Washing machine            & Ambient           & Long                      \\
Kettle            & Ambient           & Long                      \\
Vacuum cleaner            & Ambient           & Long                      \\
Dog barking            & Ambient           & Short                      \\
Bird singing            & Ambient           & Long                      \\

\end{tabular*}
\end{table}

\begin{figure*}[]
	\centering
		\includegraphics[width=17cm, height=10cm, keepaspectratio=true]{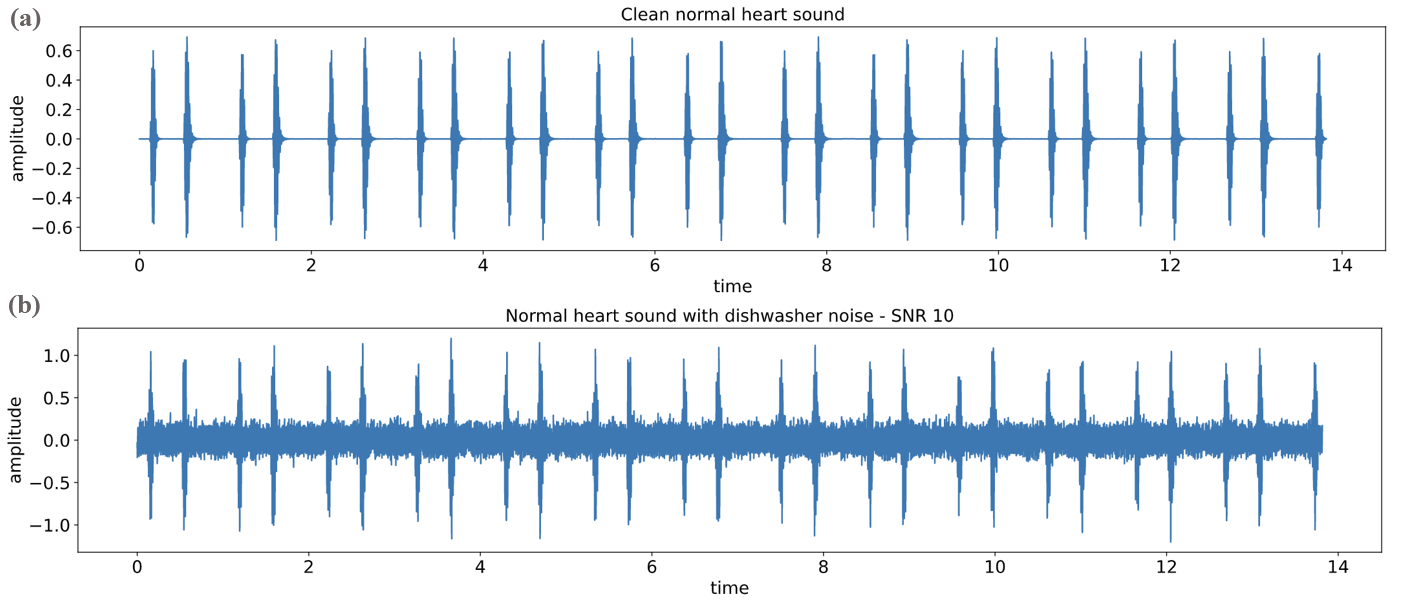}
	  \caption{(a) Phonocardiogram of a clean normal heart sound, (b) Phonocardiogram of the same heart sound contaminated with dishwasher noise}\label{fig:PCG}
\end{figure*}

\subsection{Datasets}\label{sec:datasets}

\subsubsection{Synthetic Dataset}\label{sec:datasets-syn}
This section provides the details of the synthetic heart sound dataset used in subsequent experiments and the stages of generating this dataset. Table \ref{tb:clean_hs} summarizes the specifications of the clean heart sounds, including their types and durations used to produce the synthetic dataset.

As shown in Table \ref{tb:clean_hs}, we collected multiple clean normal, and abnormal heart sounds from different resources such as publicly available datasets and YouTube. There are sixteen clean heart sounds, consisting of eight normal and eight abnormal recordings. Abnormal recordings include more common murmurs and extra heart sounds. We chose abnormalities that are more prevalent in publicly available heart sound datasets. As shown in Table \ref{tb:clean_hs}, out of these sixteen recordings, fourteen signals are over 10 seconds long, and the duration of the other two signals is 2.5-3.0 seconds. Given that we will explore the impact of heart sound duration on the accuracy of the classification models, we included both short- and long-duration signals. Short-duration recordings are long enough to include at least two heartbeat cycles, but, at the same time, they are significantly shorter than the majority of the signals. We assessed the quality of these heart sound recordings through listening and visual inspection of the waveforms to ensure they are noise-free or contain a very low noise level.

Twenty-one different noise types were mixed with each of the base clean heart sound recordings. To have a comprehensive set of noise types, we chose the noise types that are common in clinical and non-clinical environments, such as home-places. Table \ref{tb:noise_types} summarizes the details of the noise types, their groupings and durations. Noise types are categorized into four groups based on their source: color, movement, internal, and ambient. Color noises were generated through simulation, while internal and ambient noises were collected from different publicly available datasets. Movement noises were captured using a mobile phone from the body surface. Regarding the ambient noises, noise types prevalent in clinical and non-clinical environments such as homeplaces were used. Each clean heart sound was additively mixed with each noise contamination in ten different SNR levels: -10, -5, 0, 5, 10, 15, 20, 25, 30, and 40. As shown in Table \ref{tb:noise_types}, these noise types are also categorized in terms of length into short- and long-duration noises. In the case of short-duration noises, they were mixed with heart sounds at random locations. Figure \ref{fig:PCG} (a) illustrates the phonocardiogram of a clean normal heart sound, and Figure \ref{fig:PCG} (b) shows the phonocardiogram of the same heart sound contaminated with dishwasher noise where SNR is equal to 10.

Using the process described above, 3360 synthetic heart sound recordings were generated. The synthetic dataset includes 210 noisy permutations for each clean heart sound recording. The specifications of the synthetic train and test sets are as follows:

\begin{itemize}
\item Half of the samples in the synthetic dataset (1680 recordings) are placed in the train set, and the other half in the test set.
\item Train and test sets contain noisy permutations of different clean heart sound recordings: noisy permutations of heart sound numbers 1, 3, 6, 8, 9, 11, 12 and 15 are placed in the train set, while noisy permutations of heart sound numbers 2, 4, 5, 7, 10, 13, 14 and 16 are placed in the test set. The details of these base clean heart sound recordings have been provided in Table \ref{tb:clean_hs}.
\item Train and test sets are balanced across the two classes (normal and abnormal).
\item The train set contains only long-duration recordings, while the test set contains both short- and long-duration signals.
\item Noise types are the same across the train and test sets and include all twenty-one noise types, as summarized in Table \ref{tb:noise_types}.
\item SNR levels are the same across the train and test sets and include ten levels: -10, -5, 0, 5, 10, 15, 20, 25, 30, and 40.
\end{itemize}

A synthetic heart sound dataset offers several advantages over publicly available datasets for our use case. First, by synthetically adding noise to heart sound signals, we can generate recordings contaminated with a large variety of noise types common in both clinical and non-clinical settings. Also, by using different SNR levels, we can control the intensity of noise contamination in each of the recordings, allowing us to generate samples with various noise levels, from roughly clean to very noisy. Such a controlled synthetic setting enables us to thoroughly investigate the impact of different noise variables, such as noise types, groupings, intensities, and durations, on the performance of data-driven models. To date, publicly available heart sound datasets have been mainly captured in controlled environments, and recordings of such datasets are not diverse enough in terms of noise types and intensities. Also, it is very difficult to accurately measure the amount of noise in heart sound recordings in real-world datasets, and as a result, it would not be possible to provide a detailed analysis of the impact of noise and degradations on the performance of data-driven models using such datasets.

\subsubsection{PhysioNet Dataset}\label{sec:datasets-physio}
The PhysioNet heart sound dataset \cite{Liu2016open} was published as part of the PhsyioNet/Computing in Cardiology 2016 challenge. This dataset comprises six smaller subset datasets that different research groups collected across the world in controlled or uncontrolled environments. PhysioNet dataset contains 3240 heart sound recordings, out of which 2575 samples were captured from healthy subjects while 665 samples were collected from pathologic subjects. Some of the recordings were labelled as unsure, which means that they were too noisy to be categorized as normal or abnormal. For this study, we excluded these low-quality recordings from the dataset. It should be noted that this dataset does not provide any information regarding the noise content (e.g., noise type and intensity) of the recordings. In the last few years, PhysioNet dataset has been widely employed as the largest publicly available heart sound dataset to develop data-driven heart sound classification models. This dataset is used in our experiments for pre-training deep learning models.

\begin{figure*}[]
	\centering
		\includegraphics[width=18cm, height=10cm, keepaspectratio=true]{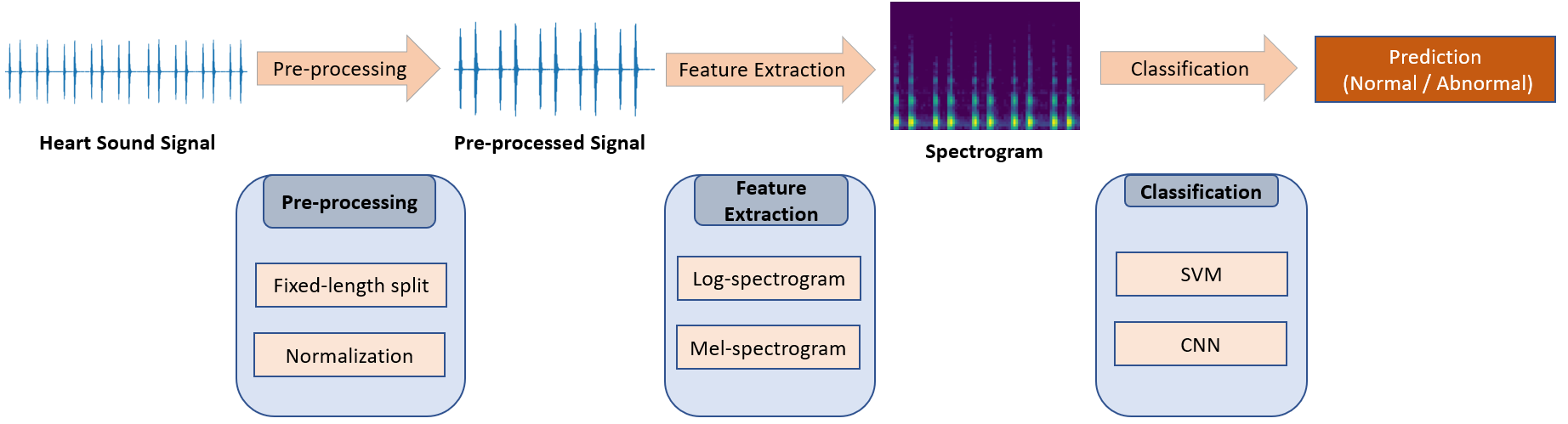}
	  \caption{Heart sound classification pipeline.}\label{fig:method}
\end{figure*}

\subsection{Data-driven Models}\label{sec:models}
As shown in Figure \ref{fig:method}, developing data-driven models for heart sound classification includes three steps: pre-processing heart sound recordings, extracting features from heart sounds, and training classification models. This section provides the details of these three stages.

\subsubsection{Pre-processing}\label{sec:models-pre}
In the pre-processing stage, long-duration heart sound recordings are split into 5- or 10-second segments and short-duration recordings are zero-padded to have similar durations. 10-second recordings are later used to develop support vector machine models, and 5-second ones are employed to develop convolutional neural network models. Given that deep learning models generally need a large number of samples for training, splitting the recordings into 5-seconds segments increases the number of available samples for training the deep learning models. Then, amplitude normalization is performed to minimize the variations in amplitudes across the signals, using the following equation (as in Ref. \cite{Chen2020Classification}):

\begin{equation}
X_{norm}(t)=\ \frac{X(t)}{Max\ (\left|X\right|)}
\end{equation}

In the above equation, $X\left(t\right)$ represents the value of the heart sound signal at the time t, and $\operatorname{Max}(|\mathrm{X}|)$ is the maximum of the absolute value of the heart sound signal.

\subsubsection{Feature Extraction}\label{sec:models-fe}
After pre-processing the recordings, Linear- and Mel-scaled Short-Time Fourier Transform (STFT) features are extracted from signals. STFT is the most widely used time-frequency feature representation for heart sound classification \cite{Bao2022Time-Frequency}. This feature representation is computed using the following equation \cite{Allen1977unified}: 
\begin{equation}
X(m, \omega)=\sum_{n=-\infty}^{\infty} x[n] \omega[n-m] e^{-i \omega n}
\end{equation}
Where $x[n]$ is the signal to be transformed and $\omega[n]$ is the window function (Hann window). After computing the STFT of the signals, spectrogram representations were computed using the following equation:
\begin{equation}
S(m, w)=|X(m, w)|^2
\end{equation}
The spectrogram gives the power of the signal for each time and frequency pair. We used Log-spectrograms as well as Mel-spectrograms as our two feature representations. Mel-spectrogram is computed by converting the linear frequency scale to the Mel scale using the following formula \cite{O'Shaughnessy2000Speech}:
\begin{equation}
\operatorname{Mel}(f)=2595 \log _{10}\left(1+\frac{f}{700}\right)
\end{equation}
We used Librosa library \cite{McFee2015librosa:} to extract the above features. Window and hop lengths were fixed at 256 and 128, respectively. As for Mel-spectrograms, 64 Mel bands were used. In order to reduce the computational cost of training the support vector machine models, the average values of the features across the time axis were computed (as in Ref. \cite{Yadav2019Machine}).

\begin{figure*}[]
	\centering
		\includegraphics[width=18cm, height=15cm, keepaspectratio=true]{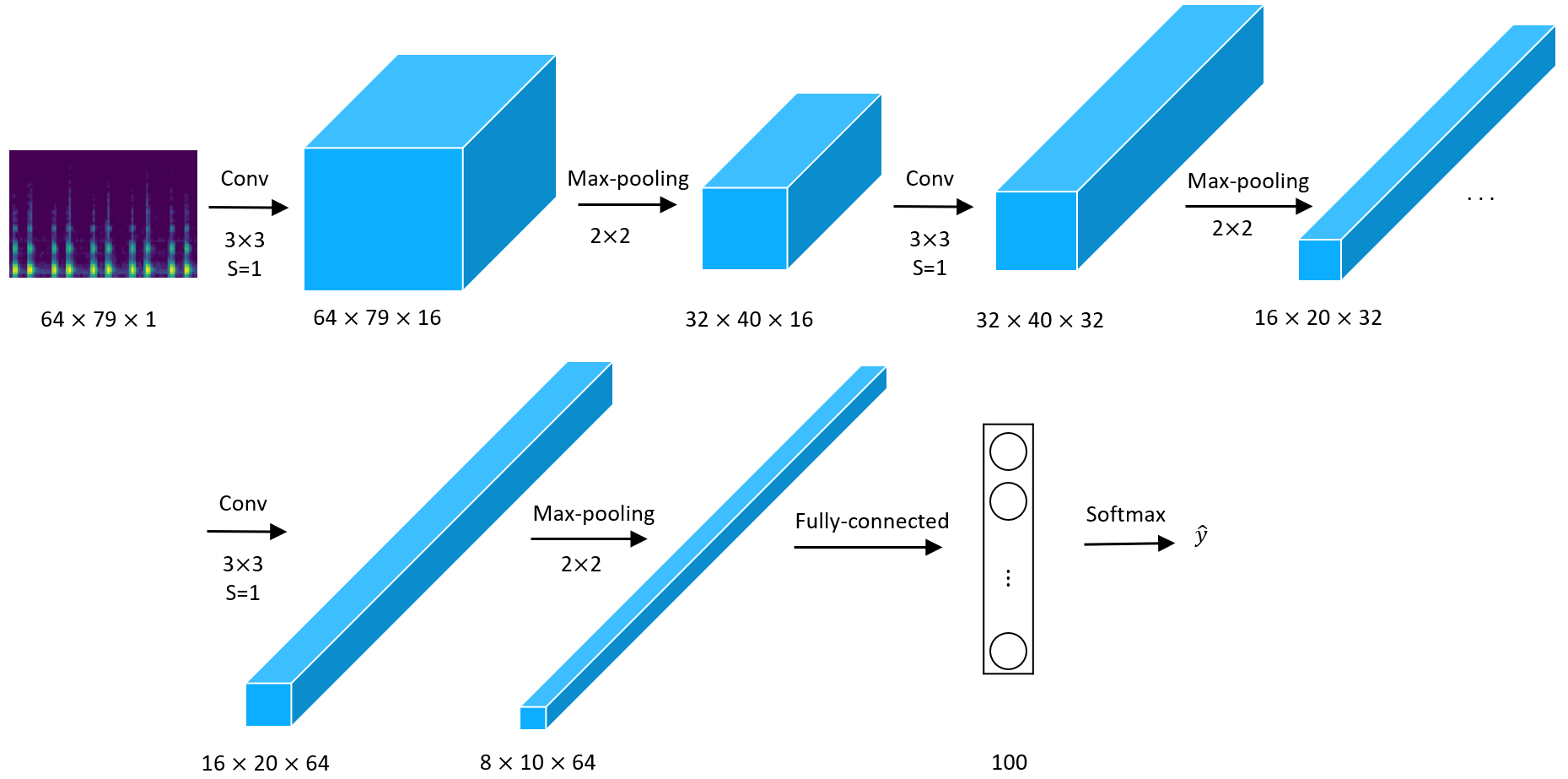}
	  \caption{Architecture of the CNN model with Mel-spectrogram as input (Mel-CNN model)}\label{fig:CNN_arch}
\end{figure*}

\subsubsection{Classification}\label{sec:models-cl}
After extracting the feature representations, they are used to train the classifiers. Two different classifiers are employed in this study: Support Vector Machine (SVM) and Convolutional Neural Network (CNN). Both classifiers have been frequently used in the field to develop heart sound classification models. Given that we use Log-spectrogram and Mel-spectrogram as input features for these classifiers, four different models are developed: a) Log-SVM, b) Mel-SVM, c) Log-CNN, and d) Mel-CNN.

For the SVM models, we implemented a linear SVM with default parameters as provided in the Scikit-Learn library \cite{PedregosaScikit-learn:}. The synthetic dataset was used to train and evaluate the SVM models.

CNN models were implemented using TensorFlow deep learning library. Figure \ref{fig:CNN_arch} shows the architecture of the CNN model with the Mel-spectrogram as input. This model consists of three convolutional layers. The first, second and third convolutional layers have 16, 32 and 64 kernels, respectively. A kernel size of (3, 3) was used for all three convolutional layers. Also, the stride was fixed at 1, and the ReLu function was used as the activation function. Each convolutional layer is followed by a max-pooling layer with a pool size of (2, 2). To reduce overfitting of the models, a dropout layer with a rate of 0.5 was used after each max-pooling layer. After convolutional and max-pooling layers, a fully connected layer with 100 neurons and the ReLu activation function was used. After this fully connected layer, another dropout layer with a rate of 0.5 is placed. The final layer of this architecture is Softmax which outputs the probability distributions of the potential outcomes (normal or abnormal).

To train the models, Adam optimization \cite{Kingma2017Adam:} with a learning rate of 0.001 and cross-entropy objective function were used. CNN models were first pre-trained on the PhysioNet dataset for 60 epochs. This way, we can ensure that the CNN models are pre-trained on a large variety of normal and abnormal heart sounds. Then, all layers except fully connected layers were frozen, and the models were fine-tuned using the synthetic training set for 10 epochs. The trained CNN models were evaluated on the synthetic test set. This process was repeated ten times, and average and standard deviation values were reported for each metric. 

It is worth noting that we first tried to train CNN models from scratch without pre-training using only the synthetic dataset. However, we observed those models were overfitting the synthetic dataset to an extreme extent. As a result, we excluded them from this study.

\subsubsection{Evaluation Metrics}\label{sec:models-metric}
The performance of the models is measured using two different metrics. The first one is recall which is used to quantify the performance of models across each class and calculated using the following formula:
\begin{equation}
\text { Recall }=\frac{TP}{TP+FN}
\end{equation}

Given that the synthetic test set used to evaluate models is balanced across normal and abnormal classes, we also use accuracy to measure the overall performance of the classification models. Overall accuracy is computed as follows:
\begin{equation}
\text { Accuracy }=\frac{TP+TN}{TP+TN+FP+FN}
\end{equation}
In the above equations, \textit{TP}, \textit{FP}, \textit{TN}, and \textit{FN} are the number of true positive, false positive, true negative, and false negative samples in the results test set, respectively.

\begin{table}[cols=4]
\caption{Performance of the classification models on the synthetic test set}\label{tb:result_cm}
\begin{tabular}{cccc}
\toprule
\multicolumn{1}{l}{Model} & \begin{tabular}[c]{@{}c@{}}Recall \\ (Normal) \%\end{tabular} & \begin{tabular}[c]{@{}c@{}}Recall \\ (Abnormal) \%\end{tabular} & \multicolumn{1}{l}{Accuracy \%} \\
\midrule
Log-SVM            & 67.9           & 73.9    & 70.9                  \\
Mel-SVM            & 68.8           & 75.6    &  72.2                 \\
Log-CNN            & 89.4 $\pm$ 1.3          & 62.6 $\pm$ 1.8  &  76.0 $\pm$ 0.4                   \\
Mel-CNN            & 82.3 $\pm$ 0.5           & 83.1 $\pm$ 1.1    &  82.7 $\pm$ 0.4                  \\
\end{tabular}
\end{table}

\begin{figure}[]
	\centering
		\includegraphics[width=8.7cm, height=5cm, keepaspectratio=true]{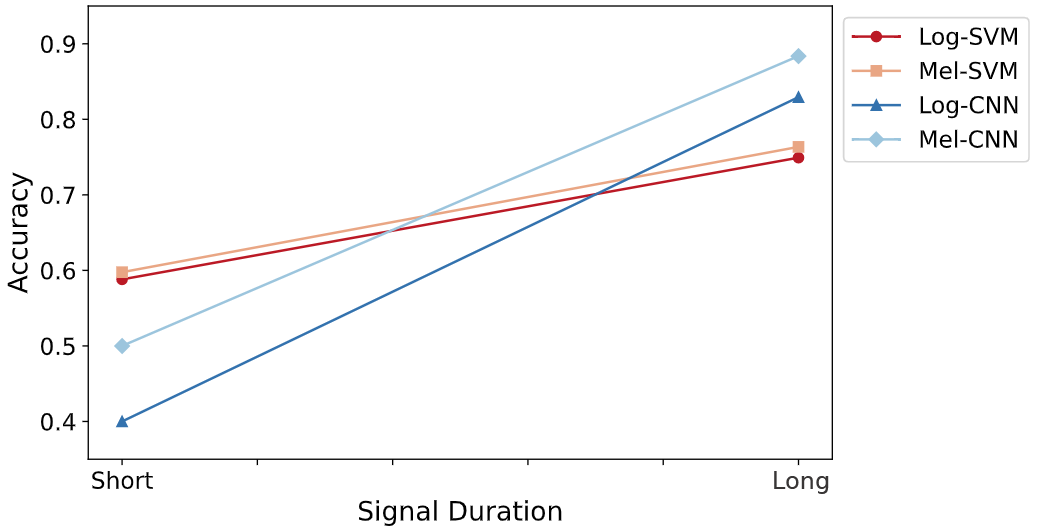}
	  \caption{Accuracy of the classification models across short- and long-duration heart sound recordings}\label{fig:sig_dur}
\end{figure}

\section{Results}\label{sec:result}

\subsection{Classification Models' Overall Performance}\label{sec:result-cmop}
Table \ref{tb:result_cm} summarizes the results of evaluating classification models on the synthetic test set.

As shown in Table \ref{tb:result_cm}, CNN models outperform SVM models in terms of overall accuracy. The Mel-CNN model achieves the highest accuracy and recall for the abnormal class, while the Log-CNN model achieves the highest recall for the normal class.

\subsection{Noise and Degradation Impact}\label{sec:result-nadi}
In this section, we explore the impact of signal duration, noise type, noise grouping, noise duration and SNR on the accuracy of the classification models.

\subsubsection{Signal Duration}\label{sec:result-nadi-sd}
As mentioned in Section \ref{sec:datasets-syn}, the synthetic dataset contains short- (2.5-3.0 seconds) and long-duration (over 10 seconds) recordings. This dataset was split into training and test sets. The training set contains only long-duration signals, while the test set used to evaluate the classification models includes both short- and long-duration recordings. Figure \ref{fig:sig_dur} depicts the impact of signal duration on the accuracy of the classification models.

As shown in Figure \ref{fig:sig_dur}, the overall accuracies of the models are considerably lower on short-duration heart sounds compared to long-duration ones. This drop in the performance of the models is more extreme for CNN models compared to SVM models. In other words, the accuracies of CNN models are over 80\% on long-duration signals, while they fall below 50\% when these models are evaluated using short-duration signals.

\begin{figure*}[]
	\centering
		\includegraphics[width=18cm, height=8cm, keepaspectratio=true]{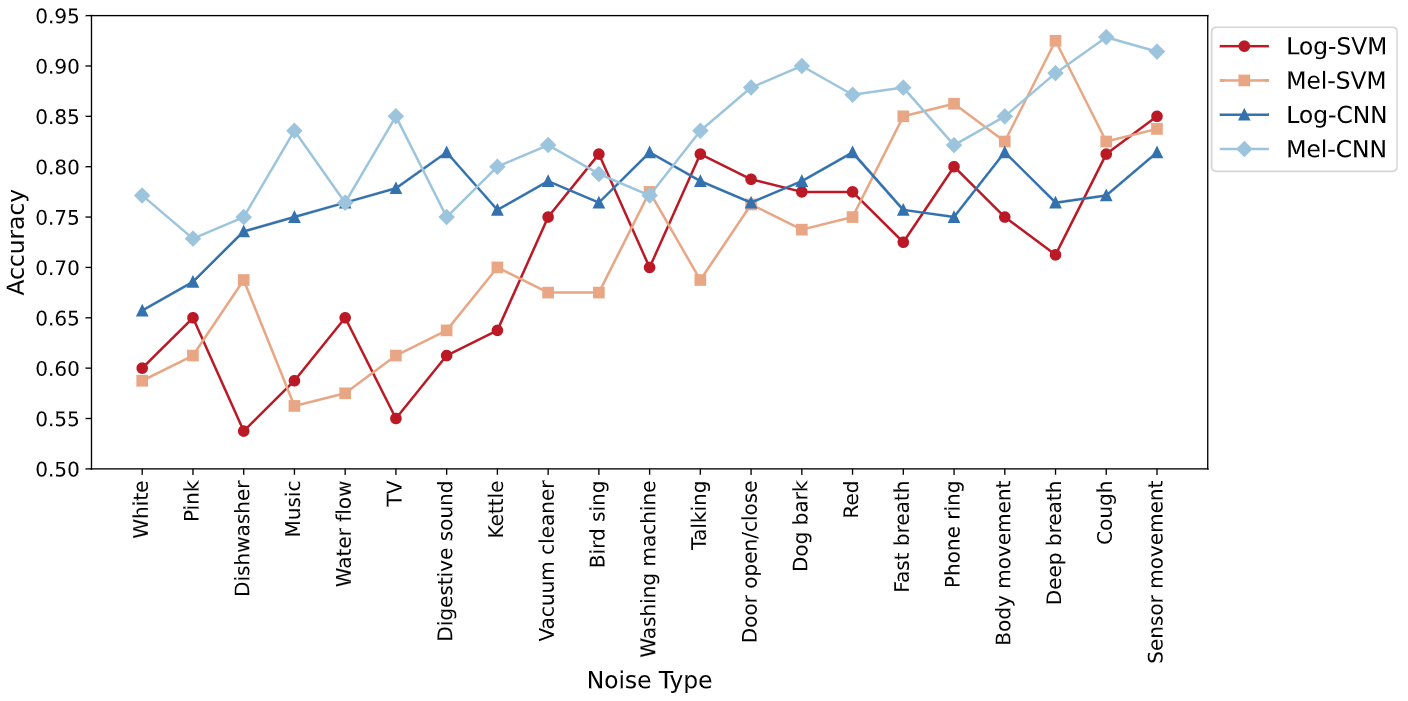}
	  \caption{Accuracy of the classification models across heart sound recordings contaminated with different noise types}\label{fig:noise_type}
\end{figure*}

\subsubsection{Noise Type}\label{sec:result-nadi-nt}
As mentioned in Section \ref{sec:datasets-syn}, the synthetic dataset contains normal and abnormal heart sounds contaminated with twenty-one noise types. Figure \ref{fig:noise_type} illustrates the overall accuracy of the models evaluated using heart sound signals contaminated with different noise types. In this plot, the noise types have been arranged based on the average accuracy of the classification models, from the noise type with the lowest (white noise) to the one with the highest (sensor movement noise) average accuracy. There are 80 samples in the test set for each noise type.

As shown in Figure \ref{fig:noise_type}, the classification models show different accuracies for heart sounds contaminated with different noise types. Also, we can observe that SVM models show larger fluctuations across different noise types than CNN models, indicating they are more sensitive to noise type than CNN models. We can see that for some noise types, like TV or dishwasher noise, SVM models offer the lowest accuracies (below 55\%), while for some others, like deep breathing and phone ring noise, they achieve much higher accuracies (over 85\%).

\begin{figure}[]
	\centering
		\includegraphics[width=8.7cm, height=5cm, keepaspectratio=true]{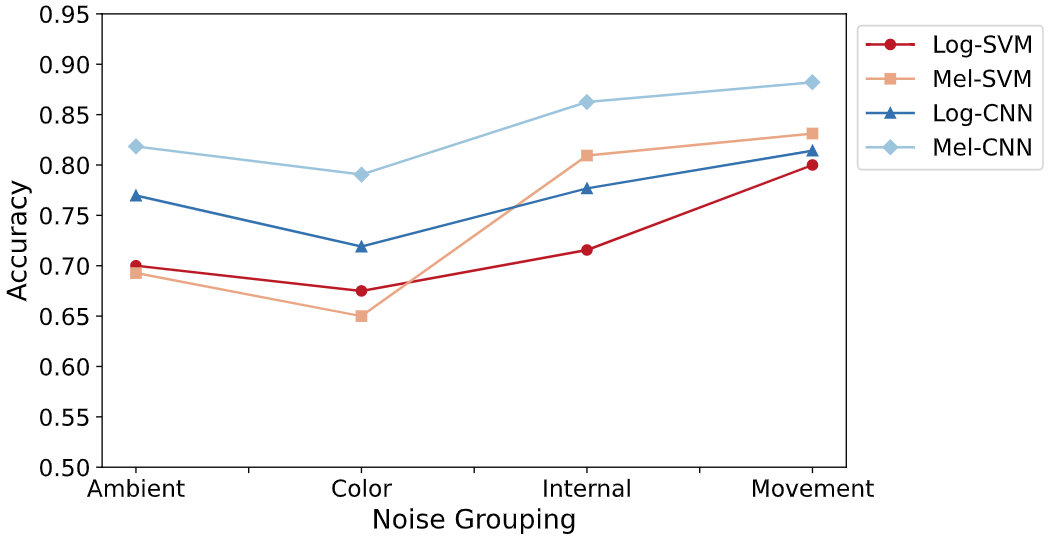}
	  \caption{Accuracy of the classification models across heart sounds contaminated with noises of four different groups}\label{fig:noise_group}
\end{figure}

\subsubsection{Noise Grouping}\label{sec:result-nadi-ng}
As summarized in Table \ref{tb:noise_types}, noise types in synthetic heart sounds are categorized into four groups: color, movement, internal and ambient.

Figure \ref{fig:noise_group} compares the accuracy of the classification models across these noise groupings. As shown in Figure \ref{fig:noise_group}, all classification models show their lowest performance on the recordings contaminated with color noises while performing best on the ones with movement noises. All models are sensitive to noise grouping, showing different accuracies across different noise groupings.

\begin{figure}[]
	\centering
		\includegraphics[width=8.7cm, height=5cm, keepaspectratio=true]{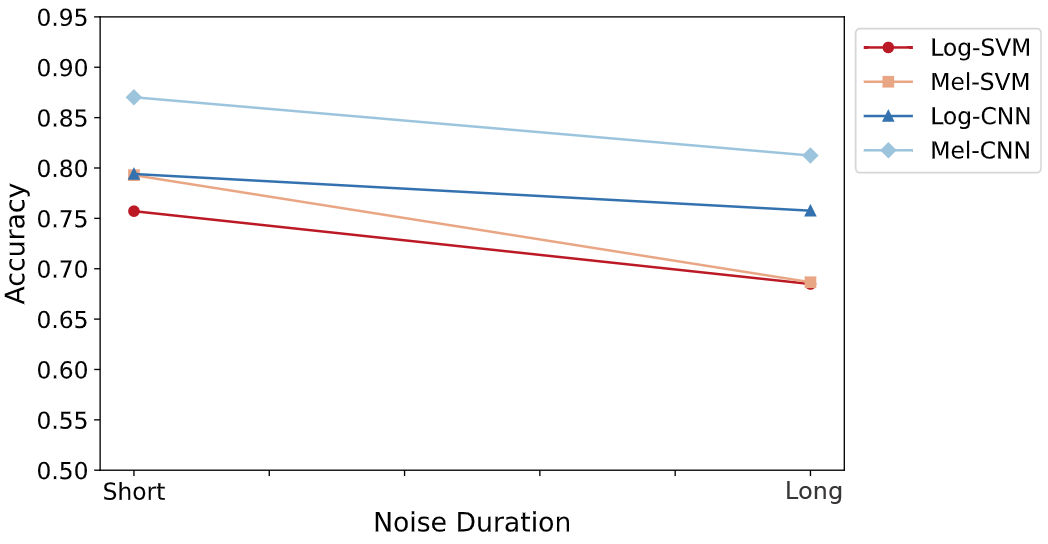}
	  \caption{Accuracy of the classification models across heart sounds contaminated with short- and long-duration noises}\label{fig:noise_dur}
\end{figure}

\subsubsection{Noise Duration}\label{sec:result-nadi-nd}
Noise contaminations used to generate the synthetic dataset can be categorized in terms of length into short- and long-duration noises (as specified in Table \ref{tb:noise_types}). Figure \ref{fig:noise_dur} compares the accuracy of the classification models across recordings contaminated with short- and long-duration noises.

As shown in Figure \ref{fig:noise_dur}, all models are sensitive to noise duration and show lower accuracies when evaluated using signals contaminated with long-duration noises compared to short-duration noises.

\begin{figure}[]
	\centering
		\includegraphics[width=8.7cm, height=5cm, keepaspectratio=true]{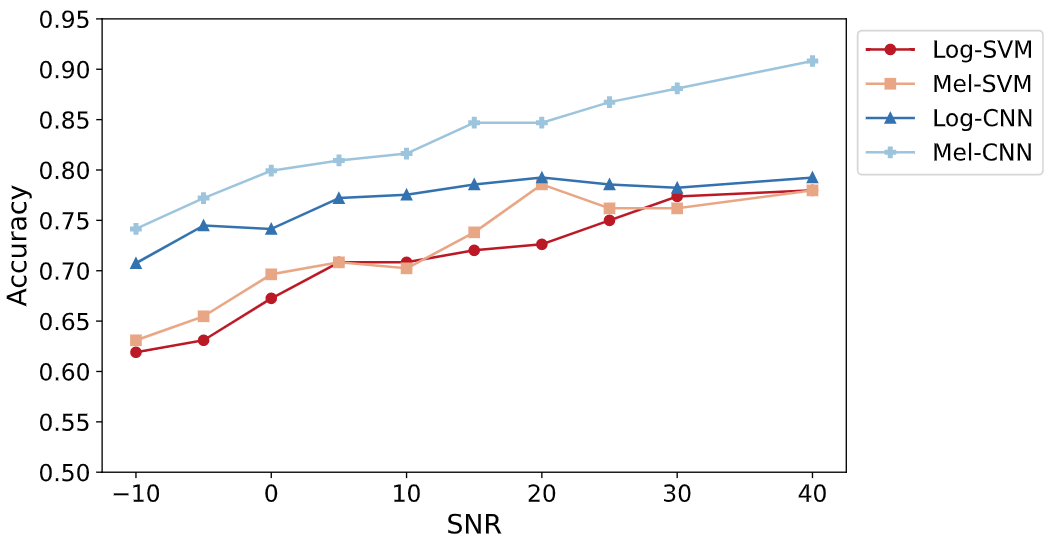}
	  \caption{Accuracy of the classification models across heart sounds with different SNR levels (from -10 to 40)}\label{fig:SNR}
\end{figure}

\subsubsection{SNR}\label{sec:result-nadi-snr}
As discussed in Section \ref{sec:datasets-syn}, in order to generate the synthetic dataset, clean heart sounds were mixed with noise contaminations with different SNR levels (from -10 to +40). For each SNR level, 168 recordings are available in the synthetic test set. Figure \ref{fig:SNR} depicts the accuracy of the classification models across heart sound recordings with different SNR levels.

As shown in Figure \ref{fig:SNR}, 
Log-SVM and Mel-CNN models show a steady increase in accuracy from SNR -10 to 40. However, in the case of Mel-SVM and Log-CNN models, we observe an increase in accuracies from SNR -10 to 20, while for SNR levels higher than 20, the accuracies are roughly unchanged. 

\begin{table}[cols=5]
\centering
\caption{Mean, standard deviation, minimum and maximum accuracies for each classification model across different noise types}\label{tb:result_noise_type}
\begin{tabular}{p{0.08\textwidth}p{0.07\textwidth}p{0.07\textwidth}p{0.07\textwidth}p{0.07\textwidth}}
\toprule
Model & Mean \% & SD \% & Min \% & Max \%  \\
\midrule
Log-SVM            & 70.9            & 9.5  &     53.8 & 85.0                \\
Mel-SVM            & 72.2            & 10.5  &     56.3 & 92.5                \\
Log-CNN            & 76.8            & 4.1  &     65.7 & 81.4                \\
Mel-CNN            & 82.9            & 5.9  &     72.9 & 92.9                \\
\end{tabular}
\end{table}

\section{Discussion}\label{sec:disc}

\subsection{Classification Models' Overall Performance}\label{sec:disc-cmop}
In Section \ref{sec:result-cmop} of the results, we presented the overall performance of the classification models. The results show that the gap between the recalls of the normal and abnormal classes is larger for the Log-CNN model compared to the other models. This can indicate that the Log-CNN model is biased towards the normal class. This bias suggests that the Log-CNN model is overfitting the synthetic dataset. 
The synthetic dataset used to train the classification models was produced using a relatively small number of base clean heart sounds. At the same time, we should note that deep learning models such as CNNs generally need a large amount of training data, making them more prone to overfitting the synthetic dataset than SVM models. The reason why we only observe this overfitting in the case of the Log-CNN model may be that the higher dimensionality of the Log-spectrogram feature representation, at twice that of the Mel-spectrogram, makes the Log-CNN model more prone to overfitting than the Mel-CNN.

\subsection{Noise and Degradation Impact}\label{sec:disc-nadi}
In Section \ref{sec:result-nadi-sd}, we discussed the impact of heart sound signal duration on the performance of classification models. All models showed a lower performance when evaluated using short-duration recordings (2.5-3.0 seconds) than long-duration signals (5-10 seconds). This finding is in line with the results of the survey we previously carried out with a group of clinicians regarding the impact of noise and degradations on the diagnosability of heart sound recordings \cite{ShariatPanah2022audio}. The majority of the survey’s respondents (92\%) stated that they needed to listen to at least six heartbeat cycles before using the recordings for diagnosis \cite{ShariatPanah2022audio}. At the same time, according to Chen et al. \cite{ChenY2020Classification}, an average heartbeat cycle is 0.8 seconds long, which means that the short-duration recordings in the synthetic dataset include around three heartbeat cycles on average. Short-duration heart sound signals, such as 1-second \cite{Alkhodari2021Convolutional,ChenY2020Classification}, or 3 seconds \cite{Xiao2020Heart,Li2021Lightweight} recordings, have been used in the field to develop data-driven heart sound classification models. However, the results show that the classification models perform considerably worse on the short-duration recordings, which suggests that short-duration signals should be avoided in situations where we can capture heart sound signals with longer durations.

In Section \ref{sec:result-nadi-nt}, we showed the performance of the classification models across heart sounds contaminated with twenty-one different noise types. Table \ref{tb:result_noise_type} summarizes the mean, standard deviation, minimum and maximum accuracies for each classificaion model across different noise types. As shown in Table \ref{tb:result_noise_type}, for all noise types, the overall accuracy of the CNN models is over 65\%, while in the case of SVM models, the accuracy goes below 57\% for a few noise types. We can also see that the standard deviations of average accuracies are larger for SVM models compared to CNN models, which confirms the lower sensitivity of CNN models to noise types. Some noise types have a more detrimental impact on the overall accuracy of the models than the other ones. For example, for SVM models, TV and dishwasher sounds are the most destructive noise types, while for CNN models, pink and white noises are more problematic. However, what is common between these classification models is that they all react to the noise type, and they do not offer the same accuracy for all noise types. We can also observe that there is only one short-duration noise type (digestive sound) in the left half of Figure \ref{fig:noise_type}, while in the right half, we can see five short-duration noises (door open/close, dog bark, body movement, cough, sensor movement). Given that in this diagram, the noise types have been ordered based on the average accuracy of the classification models, this observation confirms that the continuous long-duration noises have a more detrimental effect on the accuracy of the models than transient short-duration artifacts. These findings suggest that different noise types should not be treated the same way by quality enhancement algorithms, as each noise type influences the classification models to a different extent.

In Section \ref{sec:result-nadi-ng}, we compared the performance of the classification models on heart sounds contaminated with noises from different sources. We observed that color and ambient noises are more problematic for all models than movement and internal noises. This finding is also in agreement with the results of the survey we previously carried out with a group of clinicians. Survey respondents stated that ambient noises are more disruptive to accurate auscultation than internal or movement noises \cite{ShariatPanah2022audio}. They also stated that internal and movement noises are roughly similar in terms of their negative impact on the diagnosability of heart sounds \cite{ShariatPanah2022audio}. The observation that ambient and color noises have a more detrimental impact on the accuracy of the classification models than other noise types demonstrates that heart sounds contaminated by such noises should be prioritized over other noise sources for quality enhancement.

In Section \ref{sec:result-nadi-nd}, we explored the impact of noise duration on the accuracy of classification models. We saw that all models perform worse on heart sounds contaminated with long-duration noises than the ones contaminated with short-duration noises. These results are also aligned with the survey results from our previous study. The survey results indicated that the average quality rating for the heart sounds corrupted with long-duration noises was roughly half of those with short-duration noises \cite{ShariatPanah2022audio}. This is intuitively correct given the longer duration of distortion of the heart signal and shows that long-duration noises must be prioritized over short-duration noises for quality enhancement.

In section \ref{sec:result-nadi-snr}, we investigated the impact of the SNR of the recordings on the overall accuracy of the classification models. The results show that all classification models benefit from increased SNR levels, albeit with earlier plateauing for Mel-SVM and Log-CNN models. These results indicate that applying noise reduction techniques with the aim of improving the SNR of the recordings will have a beneficial effect on the accuracy of the classification models. However, the level of performance gain will vary from model to model.

Based on the above discussion, the major findings of this study can be summarized as follows: first, in some cases, it is possible to reduce the negative impact of noise and degradation on the classification models at the heart sound acquisition stage. For example, in Section \ref{sec:result-nadi-sd}, we observed that classification models perform worse on short-duration heart sounds compared to long-duration ones. Also, in Section \ref{sec:result-nadi-ng}, we saw that ambient noises have a more detrimental impact on classification models’ accuracy than internal or movement noises. By capturing long-duration signals or reducing ambient noises, clinicians will be able to decrease the destructive impact of such degradations on the performance of the classification models. Therefore, clinicians can use this study’s results to adapt the heart sound capture process to minimize the negative impact of such noises and degradations. Second, in order to decrease the misclassification rate of classification models, it is necessary to assess the captured heart sound signals in terms of noise and degradation characteristics. For example, in Section \ref{sec:result-nadi-snr}, we observed that classification models show a higher misclassification rate when evaluated using the signals with lower SNR levels. Also, in Section \ref{sec:result-nadi-nd}, we saw that long-duration noises have a more destructive impact on the accuracy of the classification models compared to short-duration noises. Assessing the characteristics of heart sound signals, such as SNR at the pre-processing stage, allows us to discard low-quality heart sounds or adjust the quality enhancement based on the noise characteristics of the signal. Quality enhancement of the heart sound signals has been widely employed in the field as a pre-processing step to develop heart sound classification models \cite{Gupta2007Neural,Abduh2020Classification,Chen2020Classification,Chowdhury2020Time-Frequency}. However, quality enhancement algorithms have been universally applied to heart sound recordings, irrespective of the characteristics of noise and degradation in the signal. At the same time, it has been shown that universal quality enhancement can reduce the performance of classification models \cite{Asmare2021Can}. The results of this study show that the characteristics of noise and degradation in a heart sound recording determine how and to what extent the classification models are influenced. In this regard, assessing the characteristics of the noise and degradations in heart sound signals will allow us to develop \textit{targeted} quality enhancement techniques which adapt the type and aggressiveness of quality enhancement depending on the noise content of the signals and the employed classification model.

\subsection{Limitations}\label{sec:limits}
This study has some potential limitations. Firstly, as discussed in Section \ref{sec:datasets-syn}, the synthetic heart sound dataset used to explore the impact of noises and degradations on classification models was generated using a relatively small collection of clean heart sounds. As we observed in Section \ref{sec:result-cmop}, the limited scale of the synthetic dataset is probably the reason why Log-CNN model is biased towards the normal class. Although training this model using a larger synthetic dataset could alleviate the overfitting issue, the similarity of the observed trends in the results across deep learning and SVM models indicates that the bias issue is not significant enough to impact the findings of this study. Also, this study explored the impact of noises and degradations on a small set of classification models. While a large variety of feature representations and classifiers have been employed in the field to develop heart sound classification models, in this study, we focused on the classification models used most frequently. However, there is still a slight possibility that other classification models react differently to noises and degradations in heart sound signals. Finally, as discussed in Section \ref{sec:models}, we employed a segmentation-free heart sound classification pipeline, which means that we used fixed-length signals to train and evaluate the classification models. In other words, we did not apply segmentation algorithms to segment heart sound recordings into heartbeat cycles. Therefore, the results of this study may not be generalizable to the cases where segmentation algorithms are used as one of the stages in the modelling pipeline. 

\section{Conclusion}\label{sec:conc}
Noise and degradation in heart sound recordings can reduce the accuracy of the data-driven classification models. In this study, we investigated how and to what extent different heart sound signal characteristics impact the performance of the data-driven classification models. The characteristics of heart sound signals that have been explored in this paper include signal duration, noise type, noise grouping, noise duration and SNR of the recordings. We observed that different noises and degradations do not influence the performance of heart sound classification models in the same way, which means that some are more problematic for data-driven models and some are less destructive. Clinicians can use the findings of this study to identify noise and degradations that are more problematic to classification models and consequently adapt the heart sound capture process to reduce the negative impact of those degradations. 

Universal heart sound quality enhancement, which has been frequently employed in the field as a pre-processing step, applies enhancement algorithms irrespective of the noise characteristics of the signals. However, the results of this study reinforce the importance of signal quality assessment in the heart sound classification pipelines. Quality assessment enables us to analyze the captured signals in terms of noise and degradations and limit the application of quality enhancement algorithms to specific heart sounds which do not meet a required quality threshold. The findings of this study can be leveraged to develop targeted heart sound quality enhancement approaches which adapt the type and aggressiveness of quality enhancement based on the characteristics of noise and degradations in heart sound signals.

Comparing the findings of this study with the results of the survey we previously carried out with a group of clinicians regarding the impact of noise on the diagnosability of heart sounds shows clinicians and machine learning models suffer from noise and degradations in a similar manner. We saw a strong agreement between the impact of signal duration, noise grouping and noise duration on the performance of classification models and diagnosability of heart sounds from the point of view of clinicians. Therefore, the findings from our previous survey are now backed up by the results of this study.

In future, we will extend this work by generating a larger synthetic dataset using a more diverse set of base heart sounds. Also, we will include a larger set of classification models to see if the results of this study hold for other heart sound classification models as well. The results of this study show that by analyzing the noise and degradation characteristics in heart sound signals, we will be able to develop targeted quality enhancement pipelines. As another future work, we will compare the universal and targeted heart sound quality enhancement approaches to determine which approach leads to a lower rate of misclassification in heart sound classification models.








\section*{Acknowledgement}
This work was conducted with the financial support of the Science Foundation Ireland Centre for Research Training in Digitally-Enhanced Reality (d-real) under Grant No. 18/CRT/6224. For the purpose of Open Access, the author has applied a CC BY public copyright licence to any Author Accepted Manuscript version arising from this submission.


\bibliographystyle{model1-num-names}

\bibliography{main}



\end{document}